# Optical-vortex diagnostics via Fraunhofer slit diffraction with controllable wavefront curvature

A. Bekshaev,[1*] L. Mikhaylovskaya,[1] S. Patil,[3] V. Kumar,[2] R. P. Singh[3]

[1]*Physics Research Institute, I.I. Mechnikov National University, Dvorianska 2, Odessa 65082, Ukraine*
[2]*Department of Physics, National Institute of Technology, Warangal-506004, India*
[3]*Physical Research Laboratory, Navrangpura, Ahmedabad-380009, India*
*Corresponding author: bekshaev@onu.edu.ua

**Far-field slit-diffraction of circular optical-vortex (OV) beams is efficient for measurement of the topological charge (TC) magnitude but does not reveal its sign. We show that this is because in the common diffraction schemes the diffraction plane coincides with the incident OV waist plane. Based on the examples of Laguerre-Gaussian incident beams containing a spherical wavefront component, we demonstrate that the far-field diffracted beam profile possesses an asymmetry depending on the incident wavefront curvature and the TC sign. This finding enables simple and efficient ways for the simultaneous diagnostics of the TC magnitude and sign, which can be useful in many OV applications, including the OV-assisted metrology and information processing.** © 2020 Optical Society of America

## 1. INTRODUCTION

Optical vortices (OV) are among the most interesting and attractive objects of structured light physics [1–3]. In paraxial fields, an OV appears as an isolated point of the beam cross section with zero amplitude and indeterminate phase (phase singularity); upon a round trip near this point, the field phase changes by $2\pi m$ where the integer $m$ is the topological charge (TC) of the OV. Accordingly, the beam wavefront near an OV is helical, and the OV core (zero-amplitude point) is a center for the local transverse energy circulation being the source of the orbital angular momentum (OAM) [1–4]. Due to their unique topological and singular properties, beams with OVs find many useful applications associated with the sensitive optical diagnostics and metrology [5–7], micromanipulation [8–10] and information processing [11,12].

For all fields of the OV application, rapid and reliable recognition of its rotational characteristics (determined by the magnitude and the sign of its TC) is imperative. Usually, the rich and non-trivial rotational structure of a circular OV is hidden due to its symmetry and can be revealed only in some indirect way. Standard approaches to the OV diagnostics are based on the interference with non-singular reference beams or beams with the known singular properties [1–3] but such schemes are generally complicated and cumbersome. In many situations, referenceless methods are more appropriate. For example, when a circular OV beam undergoes an astigmatic transformation, its transverse intensity distribution acquires a characteristic deformation with distinct "fingerprints" of the initial OV structure [13–15].

To the best of our knowledge, the most flexible and universal approaches exploit specific features of the OV diffraction in which the helical properties of an OV and its OAM-related circulatory nature are explicitly manifested. The simplest edge-diffraction schemes [16–18] provide spectacular demonstration of the transverse energy circulation but a reliable detection of the OV "strength" (TC magnitude $|m|$) requires additional time-consuming and precise procedures. More efficient methods enabling the "full" (TC magnitude + sign) OV diagnostics are based on the traditional approaches employing a single or double slit [19,20] and strip [21,22] Fresnel diffraction. However, the most suitable and universal means for the OV detection involve the far-field (Fraunhofer) diffraction [23–28]. The far-field scheme is, generally, less sensitive to inevitable misalignments, provides advantages of a well defined and stable reference frame as well as a considerable freedom in the choice of the registration plane, and can be easily implemented even in the ultra small-scale experimental environment. Actually, the far-field diffraction approaches are realized in the recently reported techniques adapted to the nanoscale OV diagnostics [29–32].

Despite the diversity of specific practical schemes, the interpretation of the OV-diffraction results relies on some common principles: as a rule, the immediately observable diffraction pattern (DP) contains a set of bright (dark) spots whose number is associated with the TC magnitude, and the overall pattern asymmetry indicates its sign (for example, the far-field diffraction by a triangular aperture [23–25]). However, for the most suitable cases of slit or strip diffraction, the far-field intensity pattern appears to be symmetric [24,27,28] and the "full" OV diagnostics becomes unavailable or requires additional observations.

In this paper, based on the typical example of the Laguerre-Gaussian (LG) beams [1–3], we analyze the reasons of this deficiency and propose the simple way for its elimination thus enabling the full OV diagnostics by the far-field slit (strip) diffraction. Additionally, the proposed procedure may contribute to the better visibility of informative details of the DP (e.g., its peripheral bright lobes).

## 2. THEORY

We start with a brief theoretical examination. Let a paraxial monochromatic light beam be described by the usual model with the electric field distribution expressed as $E(x,y,z) = \text{Re}[u(x,y,z)\exp(ikz - i\omega t)]$ where $\omega$ is the light frequency, $k = \omega/c$ is the wavenumber (with $c$ standing for the speed of light), and $u(x,y,z)$ is the slowly varying complex amplitude [1,2]. The beam propagates along axis $z$, and the transverse plane is parameterized by the $(x,y)$ Cartesian frame (see Fig. 1). The diffraction obstacle (slit) is situated in the plane $z = 0$, and its special role is highlighted by the special transverse coordinates' notation $(x_a, y_a)$; the slit is adjusted symmetrically with respect to the beam axis $z$.

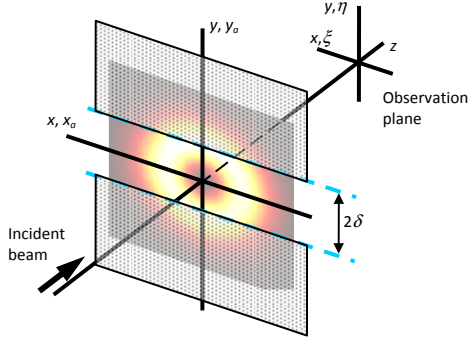

Fig. 1. Geometrical conditions of the OV diffraction.

We consider the incident $LG_{0m}$ beams with zero radial index for which the incident complex amplitude distribution in the diffraction plane is described by (see, e.g., Refs. [14,15,17,18])

$$u_{0m}(x_a, y_a, z=0) \equiv u_{0m}(x_a, y_a) = \left(\frac{x_a + i\sigma y_a}{b}\right)^{|m|} \exp\left(-\frac{x_a^2 + y_a^2}{2b^2}\right), \quad (1)$$

where $\sigma = \text{sgn}(m) = \pm 1$ is the sign of the OV TC (the winding handedness of the screw wavefront). This expression implies that the diffraction plane coincides with the incident beam waist plane (which is usual in the OV-diffraction studies [16,19–28]), and the inessential constant scaling factor depending on the beam power is omitted for simplicity; $b$ is the Gaussian envelope waist radius. Then, if the slit width equals to $2\delta$ (see Fig. 1), the DP in the observation plane is calculated via the Fresnel-Kirchhoff integral [33]

$$u_{0m}^d(x,y,z) = \frac{k}{2\pi i z}$$
$$\times \int_{-\infty}^{\infty} dx_a \int_{-\delta}^{+\delta} u_{0m}(x_a, y_a) \exp\left\{\frac{ik}{2z}\left[(x-x_a)^2 + (y-y_a)^2\right]\right\} dy_a \quad (2)$$

or, in the far-field conditions,

$$u_{0m}^d(\xi, \eta, z) = \frac{k}{2\pi i z} \exp\left(ikz\frac{\xi^2 + \eta^2}{2}\right)$$
$$\times \int_{-\infty}^{\infty} dx_a \int_{-\delta}^{+\delta} u_{0m}(x_a, y_a) \exp\left[-ik(\xi x_a + \eta y_a)\right] dy_a \quad (3)$$

which follows from (2) when $z \to \infty$. In Eq. (3), the dimensionless far-field coordinates are introduced according to relations $x/z \to k_x/k \equiv \xi$, $y/z \to k_y/k \equiv \eta$. In case of the strip diffraction, the results can be easily obtained from (2) and (3) through the Babinet principle [33].

In Fig. 2, we present the far-field intensity patterns

$$I_{0m}^d(\xi, \eta) \propto \left|u_{0m}^d(\xi, \eta, z)\right|^2, \quad (4)$$

calculated via Eq. (3) for the diffraction scheme depicted in Fig. 1 with $\delta = 0.5b$ and the incident LG beams described by Eq. (1). In full agreement with known results [27,28,32], the far-field slit-DP formed by the incident OV beam with the TC $m$ contains exactly $|m| + 1$ bright lobes, but, due to its rectangular symmetry, is quite identical for the oppositely charged OV beams. This symmetry is a direct consequence of Eqs. (1) and (3), (4) which dictate that $I_{0m}^d(\xi, -\eta) = I_{0m}^d(-\xi, \eta) = I_{0m}^d(\xi, \eta)$.

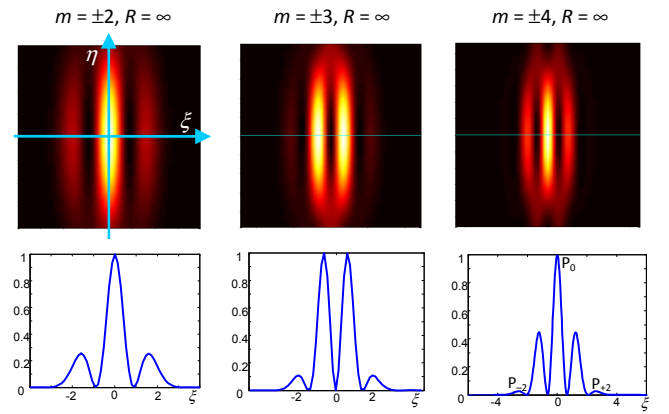

Fig. 2. (1st row) Far-field intensity patterns calculated for the incident $LG_{0m}$ beams (1) of different TCs (indicated above each column), $\delta = 0.5b$ (see Fig. 1); the far-field coordinate frame, common for all images, is indicated in the left image. (2nd row) The corresponding intensity plots (in units normalized with respect to the maximum) along the $\xi$-axis; values of $\xi$ are given in units of the divergence angle of the Gaussian envelope $\gamma = 1/(kb)$.

Note, however, that this symmetry does not hold for the Fresnel DPs determined by Eq. (2). In agreement with other similar situations [16–22], the most distinct asymmetry related to the internal energy circulation and to the sign of the incident beam TC can be observed inside the Fresnel zone where the DP is already well developed ($z > 0.5kb^2$) but the typical far-field features are not yet prevailing ($z < 2.5kb^2$); examples of Fig. 3 calculated by Eq. (2) for the center of this interval impressively illustrate this fact. So, the problem is to unite the ability of immediately detecting the TC sign, inherent in the Fresnel diffraction, with the practical advantages of the far-field scheme mentioned in the Introduction.

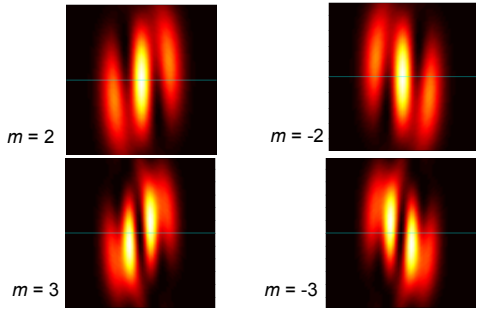

Fig. 3. DPs observed at a distance $z = 1.5kb^2$ behind the slit with $\delta = 0.5b$ for the incident LG$_{0,\pm2}$ and LG$_{0,\pm3}$ beams (1) in the diffraction scheme of Fig. 1. The far-field coordinate frame is shown in the left upper image of Fig. 2, which means that the images are presented as seen from the positive end of the $z$-axis (against the beam propagation).

It can be solved based on the known fact [34]: If the beam with initial complex amplitude distribution $u(x_a, y_a)$ produces the diffracted field described by $u(x, y, z)$ (cf. Fig. 1), the modified initial beam with the distribution

$$u_R(x_a, y_a) = u(x_a, y_a) \exp\left( ik \frac{x_a^2 + y_a^2}{2R} \right) \quad (5)$$

will produce the diffracted field described by

$$u_R(x, y, z) = \frac{1}{1 + \frac{z}{R}} \exp\left[ ik \frac{x^2 + y^2}{2(z+R)} \right] u\left( \frac{x}{1+z/R}, \frac{y}{1+z/R}, \frac{z}{1+z/R} \right). \quad (6)$$

Indeed, the transformation (5) is nothing but addition of a spherical component to the beam wavefront with preserving the same intensity profile, which can be readily performed, e.g., by usual focusing (defocusing) schemes. In turn, Eq. (6) means that the far-field ($z \to \infty$) intensity distribution created by diffraction of the modified beam (5) is proportional to $|u(R\xi, R\eta, R)|^2$, that is, reproduces (in a changed scale) the DP which could be observed with the non-modified initial beam $u(x_a, y_a)$ at a certain finite distance $z = R$ behind the screen. In application to the OV beams of Eq. (1) this implies that the DP asymmetry indicating the TC sign can be observed in the Fraunhofer plane once the diffraction plane (cf. Fig. 1) deviates from the incident beam waist plane.

Note that in case of $R < 0$ (converging beams), this procedure formally means that the far-field DP is determined by the non-modified beam behavior at $z = R < 0$, which seems non-physical. However, the Fresnel-Kirchhoff integral (2) formally is valid for any $z$; moreover, Eqs. (1), (2) and (4) dictate that

$$u_{0m}^d(x, y, -z) = \left[ u_{0,-m}^d(x, y, z) \right]^*$$

(asterisk denotes the complex conjugation), and the corresponding intensity distribution is

$$I_{0m}^d(\xi, \eta)\big|_{R<0} = I_{0,-m}^d(\xi, \eta)\big|_{R>0}.$$

Accordingly, in case of negative $R$ in Eqs. (5), (6) the DP is formed such as if the TC of the incident beam is inverted. Practically this means that both signs of $R$ are equally valid and can be chosen based on the experimental suitability.

The above conclusions are illustrated by Fig. 4 which shows examples of the far-field DPs calculated for different incident LG beams and for different values of the relative slit width $\delta/b$ and the relative wavefront curvature

$$R_s = \frac{R}{kb^2}. \quad (7)$$

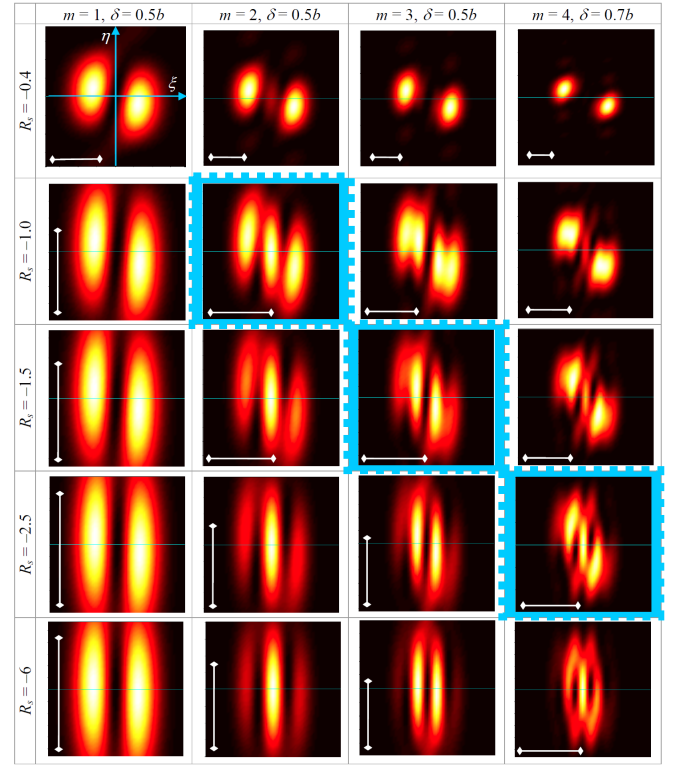

Fig. 4. Far-field slit-diffraction patterns (see Fig. 1) of converging LG$_{0m}$ beams (1) modified by (6). The TCs, slit half-widths and the wavefront curvature radii (7), accepted upon calculation, are indicated for each column and row. The far-field Cartesian coordinates are shown in the upper left image; each image is furnished with the scale bar whose length corresponds to $6\gamma = 6/(kb)$. The blue notched frames highlight the three "best" images further illustrated in Fig. 5.

It explicitly demonstrates the |m| + 1 bright lobes and, additionally, the asymmetry which indicates the OV rotational properties and the sign of its TC. When $R_s > 0$ (diverging incident beam, see Eq. (5)), the multi-lobe DP "rotates" in agreement with the incident-beam energy circulation (counter-clockwise for $m > 0$); when $R_s < 0$ (the case of Fig. 4), the rotation is opposite (remember that all the intensity images represent the patterns seen against the beam propagation, see the caption to Fig. 3).

But this is not the only remarkable feature of the diffracted OV beams with modified wavefront curvature (5). An important drawback inherent in the multi-lobe DPs of Fig. 2 is that only few central lobes are practically distinguishable. As can be seen in the second row of Fig. 2, the intensities of the peripheral lobes rapidly decay with the off-axial distance: while in case of |m| = 3 the side-lobe intensities are approximately 10% of the central maximum, for |m| = 4 the peripheral peaks $P_{+2}$ and $P_{-2}$ hardly reach 2% of the central peak $P_0$, and for higher |m| the side-lobe intensities progressively decrease. Normally, in presence of noise, this essentially restricts the maximum detectable TCs via the slit-DP. Accordingly, the "quality" of the resulting DP (its convenience for the TC diagnostics) must include not only the asymmetry but also sufficient visibility of the side lobes.

For a given incident beam [cf. Eq. (1)] this quality depends on the diffraction conditions described by parameters $R_s$ and $\delta/b$. In particular, results of the numerical analysis presented in Fig. 4 show that the slit width plays an important role, and for this reason, the value $\delta = 0.7b$, different from the previously accepted slit half-width $\delta = 0.5b$, is taken in the 4th column of Fig. 4: it appears to be more favorable for the formation of a demonstrative 5-lobe structure. As follows from Fig. 4, for |m| = 2 (2nd column of Fig. 4), the best "price-quality ratio" between the visibility of |m| + 1 lobes and the spectacular asymmetry occurs if $|R_s| = 1$; for |m| = 3 (3nd column) – if $|R_s| = 1.5$, and for |m| = 4 (4th column) – when $|R_s| = 2.5$ and $\delta = 0.7b$.

These "best" images highlighted by blue notched frames are considered in more detail in Fig. 5 where plots of the intensity distribution along the $\xi$-axis are presented. One can see that the intensities of different lobes are closer to each other than those presented in the 2nd row of Fig. 2, which is profitable for the experimental measurements. Also, the side lobes of the diffracted patterns presented in Fig. 5 are much more intense in comparison with the central lobes (for example, at |m| = 4 the peak intensities of the peripheral lobes $P_{+2}$ and $P_{-2}$ amount approximately to 16% of the central intensity $P_0$, in contrast to the right lower panel of Fig. 2 where the side-lobe intensities are ~50 times lower than the axial intensity).

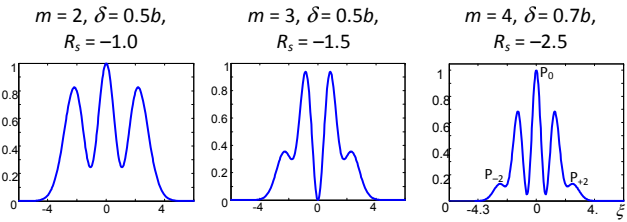

Fig. 5. The intensity distributions along the $\xi$-axis for the three "best" far-field slit-diffraction patterns of Fig. 4 (the TCs, slit half-widths and the relative wavefront curvature radii (7), accepted upon calculation, are indicated above), the horizontal scale marks are in units of $\gamma$.

## 3. EXPERIMENT

In experiment, we used a laser (iBeam smart 405-s-hp1120, Toptica Photonics) irradiating a single-mode TEM$_{00}$ beam of the wavelength $\lambda = 405$ nm ($k = 1.55 \cdot 10^5$ cm$^{-1}$). After the pin-hole spatial filtration (see Fig. 6), the nearly-Gaussian beam is transformed by the reflecting SLM (LCOS-SLM X10468, Hamamatsu Photonics) loaded with the "fork" hologram [1,2]. The SLM acts as a controllable reflecting grating which introduces an OV with the desirable TC $m$ into the first-order reflected beam. This first-order beam is selected by means of the second pin-hole filter (Fig. 6). Besides, the pin-hole diaphragm efficiently suppresses the peripheral "tails" of the SLM-generated OV beam [35] so that the good-quality doughnut-shaped mode approaches the lens L1 input (see the inset 1 in Fig. 6). After this lens (focal length $f_1 = 50$ cm), the converging OV beam is formed, whose radius and wavefront curvature vary with the $z$-distance. The slit (ThorLabs VA100, width can be adjusted with a precision of 10 μm), lens L2 and the CCD (ThorLabs BC106N-VIS, pixel size 6.45 μm and 8-bit gray level) constitute the registering unit fastened to the $z$-translation stage. As a result, the slit width and position can be adjusted to different focused-beam cross sections with desirable local beam size $b$ and the wavefront curvature radius $R$ while the lens L2 forms the Fraunhofer DP in the CCD plane.

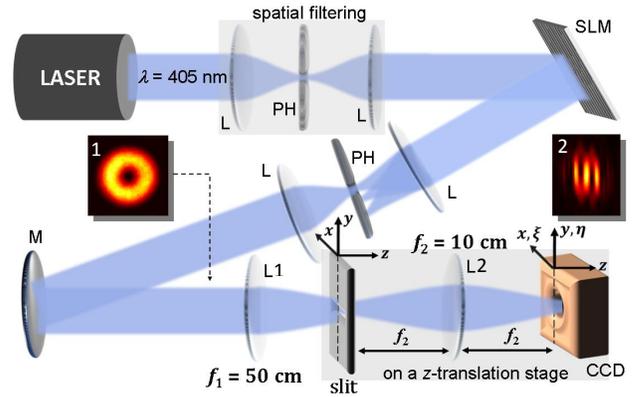

Fig. 6. Schematic of the experimental setup including (L, L1, L2) lenses, (PH) pin-hole diaphragms, (M) mirror, (SLM) spatial light modulator, and the CCD camera. Insets 1 and 2 illustrate the intensity profiles before the lens L1 and at the CCD plane, respectively. The registering unit (slit + lens L2 + CCD) can be adjusted along the longitudinal $z$-direction.

The quality of the SLM-generated OV-beams was additionally tested via registering the beam profiles at different distances after the lens L1. To this end, the slit and lens L2 were temporarily removed, and images of the focused beam were taken at different distances $z$ behind the lens L1. From the images, we found that the beams are approximately described by the LG model (1), (5) with the $z$-dependent Gaussian envelope radius obeying the standard LG-beam rule [1]

$$b(z) = b_0 \sqrt{1 + \left(\frac{z - z_0}{k b_0^2}\right)^2} \qquad (8)$$

(in contrast to Figs. 1, 3 and Eqs. (1) – (4), (6), here the plane $z = 0$ is identified with the lens L1 rather than with the slit plane). The best fitting procedure has given the waist radius $b_0 = 0.125$ mm and the waist position $z_0 \approx 60$ cm in Eq. (8) (the waist position behind the lens L1 focus is explained by the weak divergence of the incident LG beam before the lens L1).

The experimental data on the slit diffraction are slightly different from the simulations illustrated by Figs. 4 and 5. The "best" experimental DPs with articulated asymmetry and well seen side lobes are presented in Fig. 7. In this case, the slit (half-width $\delta = 0.5b$) was situated at points where the beam size equaled to $b = 0.14$ mm. The first row shows patterns registered when the slit is positioned before the waist ($z = 48$ cm): the beam converges and the wavefront curvature is negative ($R \approx -61$ cm, $R_s \approx -2$); the second row corresponds to $z = 72$ cm where $R \approx 61$ cm, $R_s \approx 2$. Despite the difference in secondary details, the experimental images show a good qualitative agreement with the theoretical prediction: the number of bright lobes (always $|m| + 1$) discloses the TC modulus whereas the overall DP asymmetry indicates the sign of $m$. What is more, relative intensities of the bright lobes in the experimental images are even better balanced than in the theoretical ones. This feature, as well as the general difference between the experimental and theoretical DPs, can be attributed, mainly, to the non-linear response of the CCD matrix due to which at the high-intensity regions, the CCD signal is saturated whereas the low-intensity regions look brighter. An additional, but minor, source of discrepancies between the simulated and experimentally obtained DPs is the deviation of the incident OV beam structure (inset 1 in Fig. 6) from the ideal LG model.

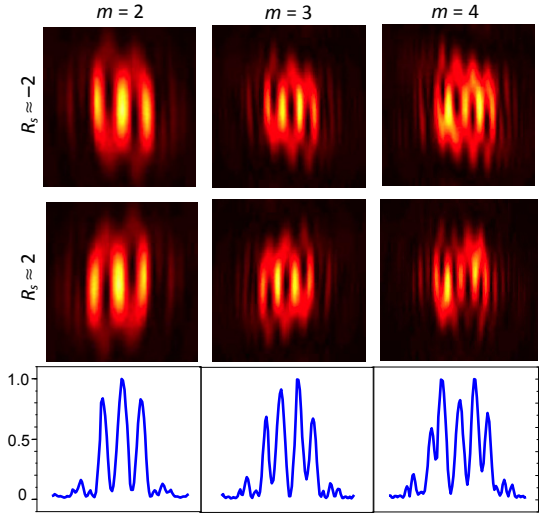

Fig. 7. Experimentally observed far-field intensity profiles of the slit-diffracted LG$_{0m}$ beams with (upper row) $R_s \approx -2$ and (middle row) $R_s \approx +2$; (bottom row) intensity distributions along the horizontal $\xi$-axis extracted from the middle-row images.

Fig. 7 shows also some extra bright fringes on both sides of the experimental DPs; however, this "ripple structure" emerging due to stray diffraction is distinctly different from the "main" lobes and practically does not deteriorate the diagnostic possibilities.

## 4. DISCUSSION AND CONCLUSION

Results of the current presentation demonstrate a new aspect of the well known phenomena associated with the slit diffraction of circular OV beams. Although the subject 'per se' is deeply investigated, in the most of known research studies the role of the incident-beam wavefront curvature eluded from the consistent analysis. In this paper, we have persuasively shown that the far-field slit-diffraction with controllable wavefront curvature can be efficient means for the "full" OV diagnostics. In addition to the known results, according to which the TC magnitude $|m|$ of the incident OV can be immediately seen from the number of bright lobes ($|m| + 1$), we demonstrate that the TC sign can also be detected due to the asymmetry of the far-field intensity distribution caused by the spherical wavefront component. If the wavefront curvature radius $R > 0$ (diverging incident beam), the multi-lobe DP looks "turned" in agreement with the energy circulation of the incident OV (counter-clockwise for positive TC when seeing against the beam propagation); if $R < 0$, the far-field DP turns oppositely. The DP asymmetry depending on the OV sign is known for the Fresnel diffraction techniques but the far-field approach provides practical advantages of a well defined and stable reference frame and is less sensitive to the system misalignments.

Additionally, our findings offer a method by which one can reproduce the Fresnel DP, characteristic for arbitrary distance behind the diffraction obstacle (slit), in the Fraunhofer (far-field) plane. As a side result, it is shown that a judicious choice of the slit width and the wavefront curvature of the incident beam may be used for optimization of the DP bright-lobes' positions and visibilities and, thus, for improvement of the OV diagnostic accuracy. The theoretical analysis is confirmed by the experimental observations.

Although the presentation is restricted to the case of slit diffraction, its ideas are equally applicable to the strip diffraction due to the Babinet's principle [33]. In this case, practically the same DPs may be masked by the strong incident-beam radiation. Nevertheless, the strip diffraction can be suitable for the OV diagnostics if the incident beam is efficiently screened by appropriate spatial filters or stops.

An analogue of such a scheme was recently realized in the nanoscale [32]. In the subwavelength situation, the vectorial nature of the optical field is essential, and the full scattering theory [33] should be applied rather than the scalar diffraction approach based on Eqs. (2), (3). However, qualitatively, the results of [32] (the scattering asymmetry observed when the incident beam is focused onto the nanowire) can be well explained by the diffraction arguments. The diffraction obstacle (nanowire) is not small compared to the longitudinal inhomogeneity of the strongly focused incident beam, so the coincidence of the waist cross section with the "diffraction plane" can only be occasional in [32]. An essential part of the incident light "meets" the obstacle in locations where the incident beam possesses a significant spherical wavefront component, which causes the DP asymmetry. In general, the results of [32] can be considered as an indirect confirmation of the principles of strip-diffraction with controllable wavefront curvature developed here.

In this paper, the standard LG$_{0m}$ model was used, but the same diagnostic principle is valid for any circular one-ring OV beam. This was numerically checked for Kummer (hypergeometric-Gaussian) beams (which are generated from incident Gaussian beams with the help of spiral phase plates or "fork" gratings [35]), and experimentally confirmed in observations with non-ideal LG beams (cf. the inset 1 in Fig. 6). For multi-ring OV beams (LG$_{pm}$ modes with non-zero radial index $p$, Bessel beams, etc. [1,2]), the slit-diffraction TC diagnostics is also available in principle but requires a special analysis because in this case the DP is formed by overlapping DPs from the separate bright rings.

In different practical situations, some details of the observed DPs (shapes of the bright lobes, their relative intensities, exact positions of the intensity maxima, etc.) can vary but the number of lobes and the DP asymmetry are still unambiguously dictated by the OV charge. In the above examples, the diagnostic features of the DPs were recognized qualitatively "by the naked eye"; however, in possible future applications, the diagnostic procedure can be suitably quantified and automated. Indeed, the number of bright lobes is easily determined from appropriate sections of the registered DP images (see, e.g., Fig. 7) whereas the asymmetry can be characterized by the "directionality factor" [32]

$$D = \frac{\int_0^\infty d\xi \int_0^\infty d\eta\, I(\xi,\eta) - \int_0^\infty d\xi \int_{-\infty}^0 d\eta\, I(\xi,\eta)}{\int_0^\infty d\xi \int_{-\infty}^\infty d\eta\, I(\xi,\eta)}, \qquad (9)$$

where $I(\xi,\eta)$ is the intensity distribution (4). Obviously, the TC sign coincides with the sign of $D$ (9) if the incident beam is divergent ($R > 0$), and is opposite if the incident beam is convergent ($R < 0$).

According to Fig. 8, the directionality magnitude distinctly depends on $|m|$, and, therefore, quantity (9) 'per se' can be used for the TC detection. However, immediate implementation of the $D$-based approaches requires further investigations of additional influences, including the system geometry (dependence $D(R)$, see Fig. 8), the incident beam deviations from the standard LG model of Eq. (1), possible noise effects, etc. This will be the subject of future work while now we consider the clear demonstrative picture, supplied by the "$|m|$ + 1 lobes" rule, to be more reliable.

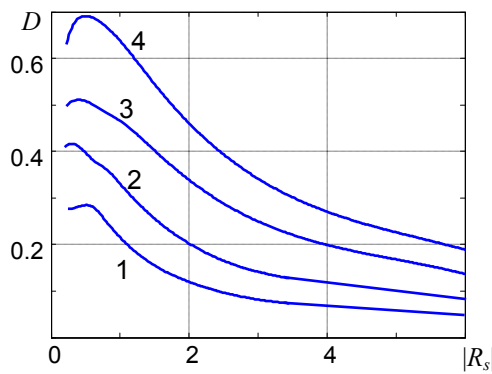

Fig. 8. Directionality (9) for the far-field patterns of Fig. 4 vs the dimensionless radius of the wavefront curvature (7); topological charges $m$ are indicated near each curve, $\delta = 0.7b$ for curve 4 and $\delta = 0.5b$ for other curves.

The modifications to the OV diagnostic techniques, proposed in this paper, can be utilized for efficient readout of the OAM states, e.g., in the OV metrology, micromanipulation techniques and for the data processing purposes [5–12]. However, it should be noted that, due to inevitable noise and limited spatial resolution, conditions for observability of the characteristic lobes becomes more and more cramped when $|m|$ grows, and for high enough $|m|$, the distinct and practically detectable pattern with exactly $|m|$ + 1 lobes is hardly achievable. This is a general deficiency of the slit(strip)-diffraction OV-diagnostic approaches [19–22,27–29], which limits their applicability by the situations with not very high $|m|$. Other methods employing the multi-lobe DP analyses [14,15,23–26] experience the same difficulties relaxed a bit due to lower difference in the lobes' intensities. Normally, this is overcome by improvement of the light-registering elements, increase in their resolving power, increase of the analyzed beam intensity, etc. As a result, cases of rather wide TC-detection ability are reported with measurable $|m|$ up to 14 [15] and even up to 100 [36] (in the latter case, under very exquisite experimental conditions hardly realizable, e.g., in practical express-diagnostics situations).

In the present work, we did not aim to reach a record but have just demonstrated a useful new principle. Anyway, the controllable wavefront curvature of the incident beam provides an additional degree of freedom which can contribute to better visibility of the whole DP structure, including the normally weak side lobes, as was noticed above (compare the images of Fig. 4 and Fig. 7 with those of Fig. 2). Further applications of this principle will surely enlarge usual facilities of the diffraction diagnostics of optical vortices.

**Funding.** Ministry of Education and Science of Ukraine (582/18, #0118U000198).

**Disclosures**. The authors declare no conflicts of interest.